\documentclass[%
 aip,
 amsmath,amssymb,
 reprint,%
]{revtex4-1}

\usepackage{graphicx}
\usepackage{dcolumn}
\usepackage{bm}

\usepackage[utf8]{inputenc}
\usepackage[T1]{fontenc}
\usepackage{mathptmx}
\usepackage{mathtools}
\global\long\def\cov{\mathcal{C}}%
\global\long\def\covll{\cov_{\{k_{x}^{(s)}\},\{k_{x}^{(i)}\}}(\lambda_{s},\lambda_{i})}%
\global\long\def\covkk{\cov_{\{\lambda_{s}\},\{\lambda_{i}\}}(k_{x}^{(s)},k_{x}^{(i)})}%
\global\long\def\covklkl{\cov(k_{x}^{(s)},\lambda_{s};\;k_{x}^{(i)},\lambda_{i})}%
\global\long\def\ket#1{|#1\rangle}%

\begin{document}

\preprint{AIP/123-QED}

\title{Fast imaging of multimode transverse-spectral correlations for twin
photons}

\author{Michał Lipka}
\email{m.lipka@cent.uw.edu.pl}
\affiliation{Centre for Quantum Optical Technologies, Centre of New Technologies, University of Warsaw, Banacha 2c, 02-097 Warsaw, Poland}

\author{Michał Parniak}
\email{m.parniak@cent.uw.edu.pl}
\affiliation{Centre for Quantum Optical Technologies, Centre of New Technologies, University of Warsaw, Banacha 2c, 02-097 Warsaw, Poland}
\affiliation{Niels Bohr Institute, University of Copanhagen, Blegdamsvej 17, 2100 Copenhagen, Denmark}

\date{\today}

\begin{abstract}
Hyperentangled photonic states – exhibiting nonclassical correlations
in several degrees of freedom – offer improved performance of quantum
optical communication and computation schemes. Experimentally, a hyperentanglement
of transverse-wavevector and spectral modes can be obtained in a straightforward
way with multimode parametric single-photon sources. Nevertheless,
experimental characterization of such states remains challenging.
Not only single-photon detection with high spatial resolution – a
single-photon camera – is required, but also a suitable mode-converter
to observe the spectral/temporal degree of freedom. We experimentally
demonstrate a measurement of a full 4-dimensional transverse-wavevector–spectral
correlations between pairs of photons produced in the non-collinear
spontaneous parametric downconversion (SPDC). Utilization of a custom
ultra-fast single-photon camera provides high resolution and a short measurement time.
\end{abstract}

\maketitle

Photonic qubits can be easily created in entangled
states, communicated over many–km distances and efficiently measured
\citep{Gisin2007}. Tremendous effort has been devoted to improving
the success rates of quantum enhanced protocols and multimode solutions, often accompanied with active multiplexing, are
one of the most promising branches of this development \citep{Collins2007,Parniak2017,mazelanik_coherent_2019,Lipka2020,Lipka2019,Wen2019,Tian2017,Himestra2019,Pu2017,Kanedaeaaw8586}, enabling both faster transfer and generation of photonic quantum states.
In particular, systems harnessing several degrees of freedom (DoF)
offer superior performance \citep{Yang2018,Graffitti2020, Brecht2015} especially
in selected protocols such as superdense coding \citep{Barreiro2008},
quantum teleportation \citep{Wang2015} or complete Bell-state analysis
\citep{Walborn2003}. Utilization of several DoFs brings a qualitatively
new possibility to create hyperentangled states exhibiting nonclassical
correlations in several DoF simultaneously with a greatly expanded
Hilbert space and informational capacity. Generation of entangled
pairs of photons in spectral, temporal, transverse wavevector, spatial,
orbital angular momentum (OAM) and with multiple DoFs has been demonstrated.
In particular spontaneous parametric down-conversion (SPDC) can be
used to generate hyperentangled states in 4 DoF simultaneously \citep{Barreiro2005}.
Nonetheless, experimental characterization of multidimensional states
remains challenging. 
Single-pixel detectors such as superconducting nanowires offer excellent timing resolution \cite{doi:10.1063/1.5010102}, as well as spectral resolution when combined with dispersive elements such as chirped fiber gratings \citep{Davis:17} or detector-integrated diffraction gratings \cite{Cheng2019}. Such setups provide a way to implement high-dimensional quantum communication \citep{Zhong_2015}, temporal super-resolved imaging \cite{Donohue2018} or observe quantum-interference in time or frequency space \cite{Jin:15,jachura_visibility-based_2018} - a promising approach for quantum fingerprinting \citep{Jachura2017,Lipka2020b}.
Single-photon-resolving cameras on the other hand naturally offer spatial or angular resolution, which can be exploited in super-resolution imaging \citep{Tenne2019,Moreau2019,Mikhalychev2019,Schwartz2013,Parniak2018}, interferometry \cite{Jachura2016}, characterization \citep{PhysRevA.99.053831,Reichert2018} or, similarly as in the previous case, observation of quantum interference effects such as in the Hong-Ou-Mandel--type experiments \citep{PhysRevX.10.031031}. Recently however, the capability of cameras has been expanded by invoking a well-known mode conversion technique, in which Sun \emph{et al.} simply observed spectral correlation with the help of a diffraction grating \citep{Sun2019}. It is thus a promising approach to use a camera to observe many DoFs simultaneously.

Here, we experimentally demonstrate a measurement of full 4-dimensional
correlations between the transverse and spectral degrees of
freedom of a twin-photon state, generated in a non-collinear type
I SPDC. An ultra-fast single-photon–sensitive camera, yielding $10^{4}$
frames per second with $100\times1952$ pixels per frame, allows to
quickly gather enormous statistic size while maintaining high resolution
due to a large number of pixels. In conjunction with recent development
in high-dimensional entanglement detection \citep{Bavaresco2018},
our single-photon detection system would enable rapid characterization
of such hyperentangled states. Furthermore, precise correlation measurements
are vital to fully utilize quantum advantage of entangled states e.g.
via non-local dispersion compensation recently demonstrated to improve
quantum key distribution rates \citep{Neumann2020}. Higher-order
correlation measurements also enable novel super-resolution imaging techniques
\citep{Mikhalychev2019,Moreau2019} which particularly benefit from
fast acquisition rates and high spatial resolution of employed single-photon
detectors.

\begin{figure*}
\includegraphics[scale=.9]{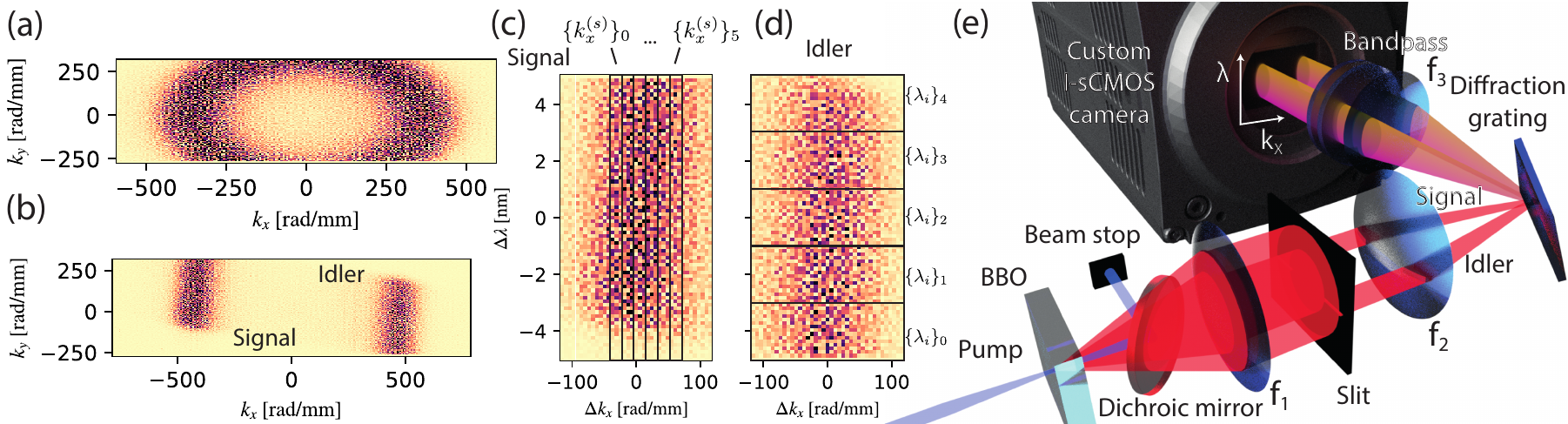}

\caption{(a) Annular ring of twin-photon emission in the far-field of a BBO
crystal mediating non-collinear type-I SPDC. Histogram of single photon
positions has been registered over $2\times10^{5}$ camera frames.
(b) Twin-photon emission in the far-field after passing
through the rectangular slit selecting $k_{y}\approx0$ and after
diffraction on the grating acting as wavelength-dependent wavevector
shift $k_{y}\rightarrow k_{y}+K_{\text{grating}}(\lambda)$. Distinct
fragments correspond to signal and idler photons, respectively. Camera
frame coordinates correspond to transverse wavevectors and wavelength
of photons. 
(c),(d) Camera regions corresponding to (c) signal and (d) idler photons.
The division into sub-regions with limited wavevectors $\{k_{x}^{(s)}\},\{k_{x}^{(i)}\}$
and wavelengths $\{\lambda_{s}\},\{\lambda_{i}\}$ has been depicted
with rectangles. For clarity only wavevector sub-regions (c) or only
wavelength sub-regions (d) has been shown. Signal and idler regions
are divided equally.
(e) Experimental setup for fast transverse-spectral correlation imaging
of twin-photons. 
\label{fig:setup}}
\end{figure*}
The employed camera prototype is an order of magnitude frame rate improvement over the off-the-shelf devices, necessary for a direct high-resolution measurement of 4-dimensional correlations. Prior approaches involved scaning the wavevector space with point detectors and used time-of-flight spectrometers for the spectral resolution, applicable at the telecom wavelengths and requiring compressed sensing techniques \citep{Montaut2018}. 
We note that a measurement in two mutually unbiased bases characterizes entanglement of pure bipartite, high-dimensional states without a state tomography \citep{Bavaresco2018}. In this context, our method would require extension to a position/time measurement basis to fully measure quantum correlations.

To generate twin-photon states we have employed a Beta Barium Borate
(BBO) non-linear crystal in the type I SPDC process with noncollinear
geometry, as depicted in Fig. \ref{fig:setup}. For the SPDC pumping
we first produce second harmonic of $70\;\text{fs}$, $800\;\text{nm}$
pulses from Ti-Sapphire laser (Spectra Physics Mai Tai, 80 MHz repetition rate) in a second, similar BBO crystal with
length $L=0.5\;\text{mm}$. The $800\;\text{nm}$ red pump is filtered-out
with dichroic mirrors and a bandpass filter ($400\;\text{nm}$, $10\;\text{nm}$
bandwidth). Blue $\lambda_{p}=400\;\text{nm}$ pump with an average
power of $70\;\text{mW}$ is focused in an $L=2\;\text{mm}$ BBO with
a Gaussian beam width of $w_{0}=70\;\mu\text{m}$ and finally filtered-out
with a dichroic mirror. The SPDC emission is far-field imaged with
a lens ($f_{1}=60\;\text{mm}$) on an adjustable rectangular slit
which selects a range of wavevectors $[-\Delta k_{y}/2,\Delta k_{y}/2]$
around $k_{y}=0$. A second lens ($f_{2}=300\;\text{mm}$) images
the BBO onto a ruled diffraction grating ($N_{\text{lines}}=1200\;\text{lines/mm}$,
resolution of $\delta\lambda=2\lambda_{p}/N_{\text{lines}}=0.66\;\text{nm}$)
mounted vertically in the Littrow configuration and at a small horizontal
angle. The grating adds a wavelength-dependent wavevector in the $y$
direction. A third lens ($f_{3}=100\;\text{mm})$ far-field images
the grating onto a single-photon camera. The effective focal size
of the setup from BBO to the camera is $f_{\text{eff}}=30\;\text{mm}$. A finite slit width $\Delta k_y$ corresponds to a resolution comparable to that of the diffraction grating $\delta \lambda$ when the image of the slit and a spectral point of $\delta \lambda$ size are compared in the camera image.

While BBO is cut for type I SPDC, by slightly adjusting
the angle of the crystal axis with respect to the pump beam, we can alter the diameter and width of the far-field annular SPDC emission ring. Comparing with theory, the crystal-axis–$z$-axis angle
(including cutting $29.2^{\circ}$) is $31.95^{\circ}\pm0.025^{\circ}$.
We estimate the overall efficiency of our at ca. $4\%$ roughly corresponding to the twin-photon generation ($50\%$), diffraction grating ($50\%$) and detection ($20\%$) efficiencies combined.
\begin{figure*}[]
\includegraphics[scale=.8]{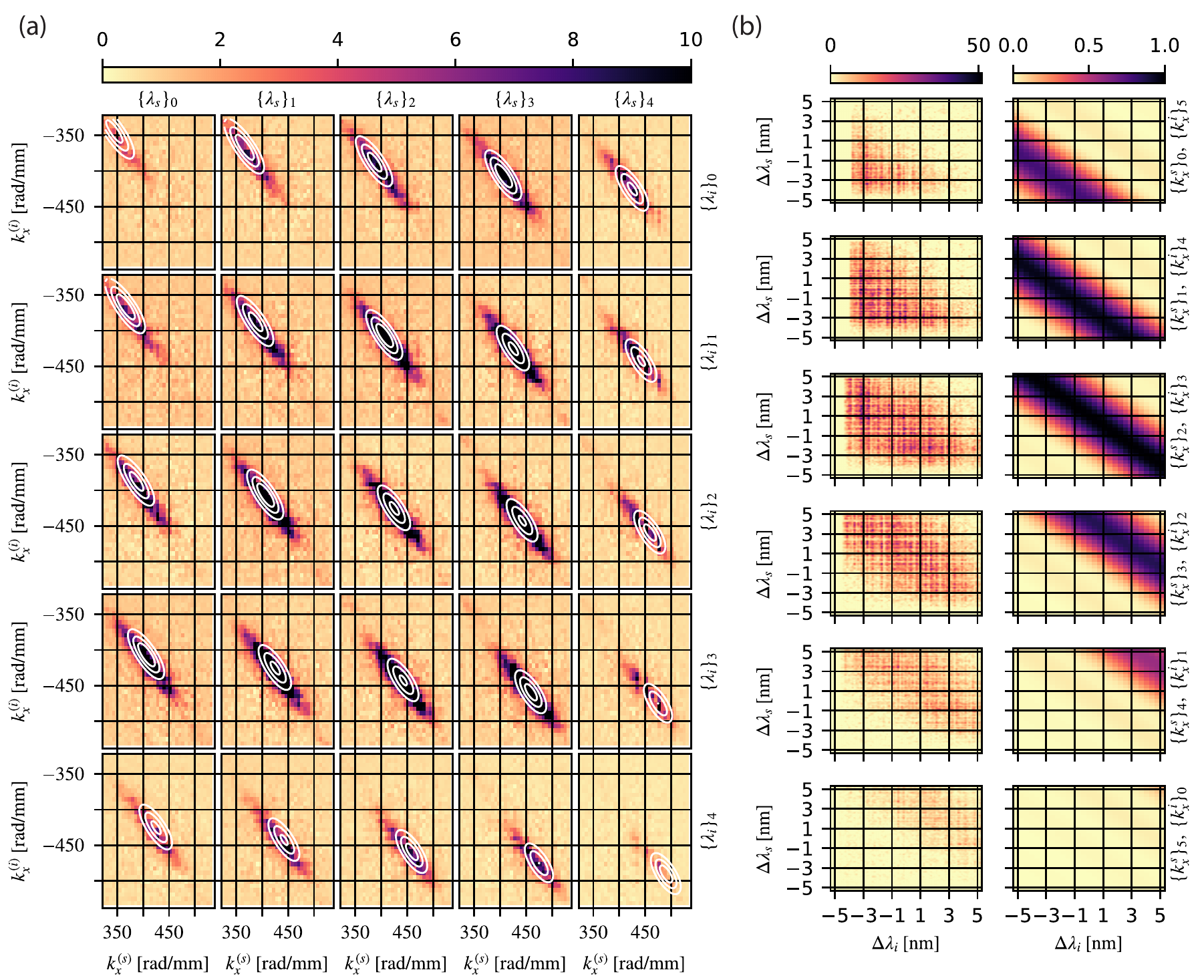}
\caption{
Correlation in joint (a) transverse wavevectors and (b) spectrum of signal\textendash idler photon pairs. (a) Color map or (b) left column represents the experimental data (photon number covariance summed over (a) spectral or (b) wavevector regions, see main text). 
(a) White contours or (b) right column represents theoretical prediction $\scriptstyle|\Psi_{\{\lambda_{s}\},\{\lambda_{i}\}}(k_{x}^{(s)},k_{x}^{(i)})|^{2}$
of a two-photon wavefunction modulus squared, (a) summed over
spectral ranges  $\scriptstyle\{\lambda_{s}\},\{\lambda_{i}\}$ or (b) transverse-wavevector ranges $\scriptstyle\{k_{x}^{(s)}\},\{k_{x}^{(i)}\}$ and normalized to a unity maximum. 
\label{fig:corkk}}
\end{figure*}

The single-photon camera consists of a two-stage image intensifier
(Hamamatsu V7090-D) with high-voltage supply (Photek
FP630) and a $10\;\text{kHz}$ gating module (Photek GM10-50B) connected
with a custom-built I-sCMOS camera based on a fast CMOS sensor (LUX2100,
pixel pitch $10\;\mu\text{m}$). Communication with the camera sensor
and low-level image processing are performed with a programmable
logic (FPGA) module (Xilinx Zynq-7020). The image processing consists
of background subtraction and single-photon localization. FPGA module is bundled with an
ARM-family processor, providing Ethernet data transfer to PC.
The CMOS sensor is set for lowest time-dependent noise
at the cost of lower dynamic range. With a faster
gating module the camera could operate at $10^5$ frames
per second with a frame of $10\times1952$ px. 
Single-photon sensitivity is achieved by operating the image intensifier
(II) in the Geiger mode (on/off) \citep{Lipka2018} (see supplementary material - SM).

The gating time was $1.2\;\mu\text{s}$ with an average
of $\bar{n}_{tot}=0.12$ photons per frame.
Signal and idler photons are observed in $40\times70\;\text{px}$ regions each corresponding
to $416\;\text{rad/mm}\times5.1\;\text{nm}$ (with $5.95\;\text{rad}/(\text{mm}\times\text{px})$
and $0.127\;\text{nm}/\text{px}$). The Gaussian mode size $\sigma_{k\text{-mode}}$
was predicted to be $6.2\;\text{rad/mm}$ and measured as $7.1\pm0.3\;\text{rad}/\text{mm}$. The spectral mode size was measured to
be $\sigma_{\lambda\text{-mode}}=4.20\pm0.06\;\text{nm}$. We define the mode
sizes as the Gaussian widths of a two-dimensional second-order photon number correlation in the sum coordinates 
$(k_x^{(s)}+k_x^{(i)}, \lambda_s+\lambda_i)$ (see SM).
Using the theoretical prediction of the joint wavefunction we numerically
get $M=1/\sum_{j=0}^{\infty}\lambda_{j}^{2}\approx4.7$ accessible entangled modes, where
$\lambda_{j}$ are the Schmidt coefficients (see SM). Note that for our considerations regarding the mode size and the number of modes we implicitly assumed a Gaussian two-photon wavefunction (leading to Gaussian second-order correlations) and as well as purity of the generated state.

While with a spectrally broad, focused pump beam and a
short crystal, the SPDC emission is highly multimode in the spectral
and transverse DoF, we begin with a single pair of signal ($s$)–
idler ($i$) modes. A two-mode squeezed state $\ket{\psi}=\sum_{j=0}\chi^{j/2}\ket j_{s}\ket j_{i}$,
generated in SPDC can be approximated to the first order in $\sqrt{\chi}$
as a pair of photons $\ket 1_{s}\ket 1_{i}$. Consider the
joint wavefunction in transverse-wavevector and spectral coordinates: 
\begin{equation}
\Psi(\boldsymbol{k}_{s,\perp},\lambda_{s};\boldsymbol{k}_{i,\perp},\lambda_{i})=\langle\boldsymbol{k}_{s,\perp},\lambda_{s}|1\rangle_{s}\langle\boldsymbol{k}_{i,\perp},\lambda_{i}|1\rangle_{i}.
\end{equation}
We directly measure the $x$ component of the transverse
wavevector, while selecting photons with $k_{y}\approx0$.
Before measurement, a diffraction grating maps the spectral DoF onto
$k_{y}(\omega)$. Single-photon camera detects the number of photons
with a given transverse–spectral coordinate $n(k_{x}^{(\xi)},\lambda_{\xi})\in\{0,1\};\;\xi\in\{s,i\}$
separately in signal and idler arms. With a large number of observed
frames, the average over frames $\scriptstyle\langle n(k_{x}^{(\xi)},\lambda_{\xi})\rangle$
gives an estimate for the probability of detecting a photon at given
coordinates. Hence, the photon number covariance:
\begin{multline}
\covklkl=\\
\langle n(k_{x}^{(s)},\lambda_{s})n(k_{x}^{(i)},\lambda_{i})\rangle-\langle n(k_{x}^{(s)},\lambda_{s})\rangle\langle n(k_{x}^{(i)},\lambda_{i})\rangle\label{eq:covklkl}
\end{multline} 
estimates the probability of detecting a non-accidental coincidence – pair
of correlated signal and idler photons in a single camera frame –
with given spectral and transverse coordinates, modeled by $|\Psi(\boldsymbol{k}_{s,\perp},\lambda_{s};\boldsymbol{k}_{i,\perp},\lambda_{i})|^{2}$.
The camera gating time encompasses ca. 96 pump laser repetitions, hence 96 temporal modes are aggregated in each camera frame, producing accidental coincidences between photons from different temporal modes. The second term in Eq. (\ref{eq:covklkl}) roughly corresponds to these accidental coincidences.
For visualization, we sum the covariance over selected sub-regions
either in wavelengths $\{\lambda_{s}\},\{\lambda_{i}\}$ or wavevectors
$\{k_{x}^{(s)}\},\{k_{x}^{(i)}\}$ yielding:
\begin{align}
\covkk&=
\smashoperator[lr]{\sum_{\substack{\lambda_{s}\in\{\lambda_{s}\} , \lambda_{i}\in\{\lambda_{i}\}}}}\covklkl,\label{eq:covkk}\\
\covll&=
\smashoperator[lr]{\sum_{\substack{k_{x}^{(s)}\in\{k_{x}^{(s)}\} ,k_{x}^{(i)}\in\{k_{x}^{(i)}\}}}}\covklkl.\label{eq:covll}
\end{align}
The selected sub-regions are depicted in Fig. \ref{fig:setup} (c),(d)
on a histogram of signal and idler positions in wavevector–wavelength
space. During the measurement we gathered $10^{9}$ camera frames,
each serving as a separate experiment repetition.
The joint covariance in transverse wavevector coordinates $\covkk$
is depicted in Fig. \ref{fig:corkk} (a) with each panel corresponding
to a different pair of wavelength sub-regions $\{\lambda_{s}\},\{\lambda_{i}\}$.
Good agreement with the theoretical prediction can
be observed with only the crystal-axis--z-axis angle fitted (a small deviation from the cutting angle). The details of wavefunction calculation can be found
in Supplementary Material.
Similarly, the joint covariance in spectral coordinates $\covll$
is depicted in Fig. \ref{fig:corkk} (b) for selected pairs of sub-regions
in which the covariance is non-vanishing. 
selecting $k_y\approx0$ limits observations of the 6-dimensional space of two-photon transverse-spectral correlations to a 4-dimensional slice. The limitation can be relevant for complex transverse correlations e.g. from a biaxial crystal. 

We have demonstrated a capability to measure 4 dimensional transverse-wavevector–spectral
correlations between pairs of photons generated in non-collinear
SPDC. Due to a custom single-photon camera with very fast acquisition
rates (an order of magnitude improvement) we were able to gather statistics
of $10^{9}$ camera frames (experiment repetitions) in roughly one
day. Large statistics enabled faithful reconstruction of bi-photon
wavevefunction in spectral and transverse-wavevector coordinates.
For this demonstration we selected a single component of the transverse
wavevector which is far-field imaged (mapped) onto positions on the
camera frame, similarly the spectral part is mapped onto positions
with a diffraction grating. Importantly, our system is inherently
multimode and can be adapted for measurements in different mode bases
e.g. orbital angular momentum and for different degrees of freedom.

\paragraph*{Funding}

Ministry of Science and Higher Education (DI2018 010848); Foundation
for Polish Science (MAB/2018/4 ``Quantum Optical Technologies''); Office of Naval Research (N62909-19-1-2127).

\paragraph*{Acknowledgements}

The \textquotedbl Quantum Optical Technologies” project is carried
out within the International Research Agendas programme of the Foundation
for Polish Science co-financed by the European Union under the European
Regional Development Fund. We would like to thank W. Wasilewski
for fruitful discussions and K. Banaszek for the generous support.

\paragraph*{Disclosures}

The authors declare no conflicts of interest.

\bibliographystyle{apsrev4-1}
\bibliography{4cor}

\clearpage
\part*{Supplementary Material}

\paragraph*{Bi-photon amplitude and photon number covariance}

Let us begin with the positive part of the blue pump classical electric
field:

\begin{multline}
E_{p}^{(+)}(\boldsymbol{r},t)=\\
\mathcal{E}_{p}\int\mathrm{d}^{2}\boldsymbol{k}_{p,\perp}\mathrm{d}\omega_{p}A_{p}(\boldsymbol{k}_{p,\perp},\omega_{p})\exp[i(\boldsymbol{k}_{p}\cdot\boldsymbol{r}-\omega_{p}t)],
\end{multline}
where $\mathcal{E}_{p}$ denotes the pump pulse amplitude, $\boldsymbol{k}_{p,\perp}$
its transverse wavevector and $A_{p}$ corresponds to the normalized
slowly varying envelope of the pulse. From the wavefunction of the
state generated in SPDC we shall consider only the biphoton part \citep{Kolenderski2009},
which can be denoted as:

\begin{multline}
|\Psi\rangle=\int\mathrm{d}^{2}\boldsymbol{k}_{s,\perp}\mathrm{d}^{2}\boldsymbol{k}_{i,\perp}\mathrm{d}\omega_{s}\mathrm{d}\omega_{i}\Psi(\boldsymbol{k}_{s,\perp},\omega_{s};\boldsymbol{k}_{i,\perp},\omega_{i})\times\\
\hat{a}^{\dagger}(\boldsymbol{k}_{s,\perp},\omega_{s})\hat{a}^{\dagger}(\boldsymbol{k_{i,\perp}},\omega_{i})|\text{vac}\rangle,
\end{multline}
where $s,i$ indices correspond to signal and idler photons, respectively.
For a crystal with length $L$ and assuming the $z$ axis points along
the pump beam's central wavevector, the biphoton amplitude $\Psi(\boldsymbol{k}_{s,\perp},\omega_{s};\boldsymbol{k}_{i,\perp},\omega_{i})$
is given by:

\begin{multline}
\Psi(\boldsymbol{k}_{s,\perp},\omega_{s};\boldsymbol{k}_{i,\perp},\omega_{i})=\mathcal{N}\int_{-L/2}^{L/2}\mathrm{d}z\:\big\{\\
A_{p}(\boldsymbol{k}_{s,\perp}+\boldsymbol{k}_{i,\perp},\omega_{s}+\omega_{i})\exp[i\Delta k_{z}(\boldsymbol{k}_{s,\perp},\omega_{s};\boldsymbol{k}_{i,\perp},\omega_{i})z]\big\},
\end{multline}
where the phase mismatch $\Delta k_{z}(\boldsymbol{k}_{s,\perp},\omega_{s};\boldsymbol{k}_{i,\perp},\omega_{i})$
is determined by the $z$ components of the wavevectors:
\begin{multline}
\Delta k_{z}(\boldsymbol{k}_{s,\perp},\omega_{s};\boldsymbol{k}_{i,\perp},\omega_{i})=k_{p,z}(\boldsymbol{k}_{s,\perp}+\boldsymbol{k}_{i,\perp},\omega_{s}+\omega_{i})-\\
k_{s,z}(\boldsymbol{k}_{s,\perp},\omega_{s})-k_{i,z}(\boldsymbol{k}_{i,\perp},\omega_{i}).
\label{eq:delkz}
\end{multline}

Note that we work in the paraxial approximation. In particular since $|\boldsymbol{k}_\perp|\ll|\boldsymbol{k}|$ and the crystal's index of refraction changes slowly over the range of observed wavevectors, we can assume that the transverse wavevector components within and outside the crystal are equal (at the crystal-air boundary the emission angle increases by a factor equal to the crystal's index of refraction but so does the total wavevector, hence the transverse component must remain unchanged in the small-angle approximation).

Integrating Eq. (\ref{eq:delkz}) along $z$ we get:
\begin{multline}
\Psi(\boldsymbol{k}_{s,\perp},\omega_{s};\boldsymbol{k}_{i,\perp},\omega_{i})=\\
\mathcal{N}A_{p}(\boldsymbol{k}_{s,\perp}+\boldsymbol{k}_{i,\perp},\omega_{s}+\omega_{i})\text{sinc}[\frac{L\Delta k_{z}(\boldsymbol{k}_{s,\perp},\omega_{s};\boldsymbol{k}_{i,\perp},\omega_{i})}{2}].\label{eq:psi}
\end{multline}
In our experiment we select a narrow range of $k_{y}$ around $k_{y}=0$.
Hence, we shall simplify the notation $\mathbf{k_{\alpha,\perp}}\rightarrow k_{x}^{(\alpha)};\;\alpha\in\{s,i\}$.
Importantly, $|\Psi(k_{x}^{(s)},\omega_{s};\;k_{x}^{(i)},\omega_{i})|^{2}$
is proportional to the probability of simultaneously generating a
signal photon with transverse wavevector $k_{x}^{(s)}$ and wavelength
$\lambda_{s}=2\pi c/\omega_{s}$ and an idler photon with $k_{x}^{(i)}$
and $\lambda_{i}=2\pi c/\omega_{i}$.
Clearly, if the bi-photon term
vanishes $|\Psi(k_{x}^{(s)},\omega_{s};\;k_{x}^{(i)},\omega_{i})|^{2}=0$
for some coordinates $(k_{x}^{(s)},\omega_{s},k_{x}^{(i)},\omega_{i})$,
the probability of observing a photon pair in $(k_{x}^{(s)},\omega_{s},k_{x}^{(i)},\omega_{i})$
is equal to the product of marginal probabilities of observing a signal
photon at $(k_{x}^{(s)},\omega_{s})$ and of observing an idler photon
at $(k_{x}^{(i)},\omega_{i})$. Hence, $|\Psi(k_{x}^{(s)},\omega_{s};\;k_{x}^{(i)},\omega_{i})|^{2}$
is proportional to photon number covariance between signal and idler
modes $\covklkl$.

For direct comparison with measured covariances we sum the modulus
squared amplitudes in the selected wavelength and wavevectors regions,
respectively:

\begin{multline}
|\Psi_{\{\lambda_{s}\},\{\lambda_{i}\}}(k_{x}^{(s)},k_{s}^{(i)})|^{2}=\\
\sum_{\lambda_{s}\in\{\lambda_{s}\},\lambda_{i}\in\{\lambda_{i}\}}|\Psi(k_{x}^{(s)},2\pi c/\lambda_{s};\;k_{x}^{(i)},2\pi c/\lambda_{i})|^{2},
\end{multline}
\begin{multline}
|\Psi_{\{k_{x}^{(s)}\},\{k_{x}^{(i)}\}}(\lambda_{s},\lambda_{i})|^{2}=\\
\sum_{k_{x}^{(s)}\in\{k_{x}^{(s)}\},k_{x}^{(i)}\in\{k_{x}^{(i)}\}}|\Psi(k_{x}^{(s)},2\pi c/\lambda_{s};\;k_{x}^{(i)},2\pi c/\lambda_{i})|^{2}.
\end{multline}

\paragraph*{Non-classical correlations and mode size}

\begin{figure}
\includegraphics[width=1\columnwidth]{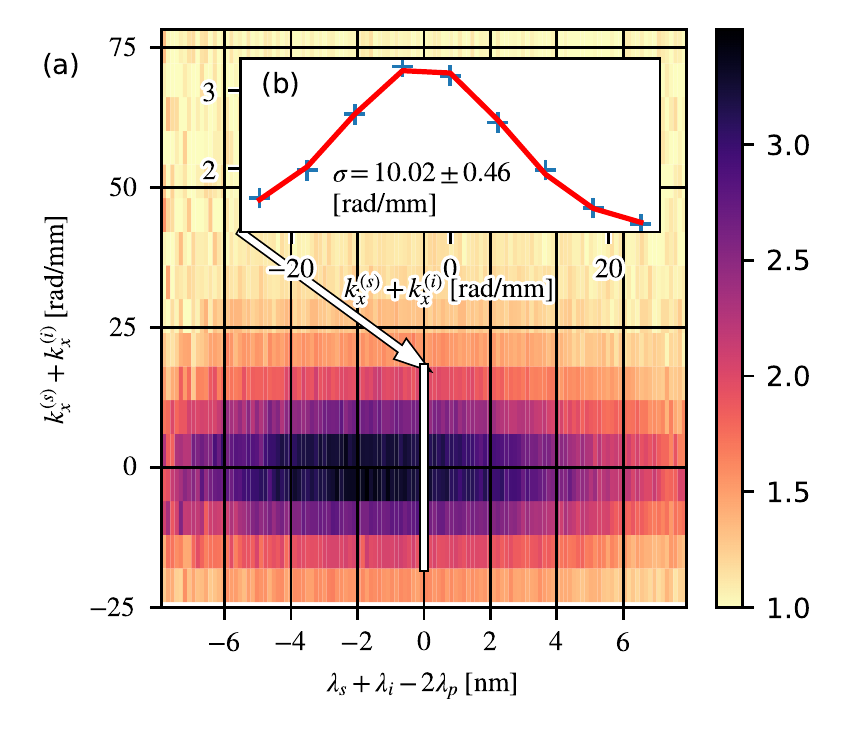}

\caption{(a) Second order photon number correlation function $g^{(2)}(k_{x}^{(s)}+k_{x}^{(i)},\lambda_{i}+\lambda_{s})$
between signal and idler photons, in sum coordinates of spectral and
temporal DoF. (b) Cross section for $\lambda_{i}+\lambda_{s}=2\lambda_{p}$.
Data (blue crosses) is depicted against a Gaussian fit (red solid)
with Gaussian width parameter $\sigma=10.02\pm0.46\;[\text{rad/mm}]$.\label{fig:g2}}
\end{figure}
To quantify the non-classical character of signal–idler correlations and estimate the mode size
we employ the second-order photon number correlation function defined
for a single pair of signal–idler modes as:
\begin{multline}
g^{(2)}(k_{x}^{(s)},\lambda_{s};\;k_{x}^{(i)},\lambda_{i})=
\frac{\langle n(k_{x}^{(s)},\lambda_{s})n(k_{x}^{(i)},\lambda_{i})\rangle}{\langle n(k_{x}^{(s)},\lambda_{s})\rangle\langle n(k_{x}^{(i)},\lambda_{i})\rangle}.\label{eq:g2full-1}
\end{multline}
For visualization we sum the coincidences and the normalizing factor
over the uncorrelated directions to get:

\begin{multline}
g^{(2)}(k_{+},\lambda_{+})=\\
\sum_{k_{-},\lambda_{-}}\langle n(k_{x}^{(s)},\lambda_{s})n(k_{x}^{(i)},\lambda_{i})\rangle(k_{+},k_{-};\;\lambda_{+},\lambda_{-})/\\
\big\{\sum_{k_{-},\lambda_{-}}\langle n(k_{x}^{(s)},\lambda_{s})\rangle(k_{+},k_{-};\;\lambda_{+},\lambda_{-})\\
\times\langle n(k_{x}^{(i)},\lambda_{i})\rangle(k_{+},k_{-};\;\lambda_{+},\lambda_{-})\big\}
\end{multline}
with $k_{\pm}=k_{x}^{(s)}\pm k_{i}^{(i)}$, $\lambda_{\pm}=\lambda_{s}\pm\lambda_{i}$
and where we implicitly transformed the mean photon numbers to $\pm$
coordinates. 
As depicted in Fig. \ref{fig:g2}, the cross-section for degenerate wavelength $\lambda_{i}+\lambda_{s}=2\lambda_{p}$
has a Gaussian shape with width $\sigma=10.02\pm0.46\;[\text{rad/mm}]$.
This width corresponds to the the transverse wavevector mode size
of SPDC emission as 
\begin{equation}
\sigma_{k\text{-mode}}=\frac{\sigma}{\sqrt{2}},    
\end{equation} 
where
the $\sqrt{2}$ factor comes from the Jacobian of $(k_{x}^{(s)},k_{x}^{(i)})\rightarrow(k_{x}^{(s)}+k_{x}^{(i)},k_{x}^{(s)}-k_{x}^{(i)})$
transformation. Similarly, the Gaussian fit for $k_{x}^{(s)}=-k_{x}^{(i)}$
cross-section gives $\sigma_{\lambda\text{-mode}}=4.20\pm0.06\;\text{nm}$.
If we trace over the idler (signal) mode, the remaining signal (idler)
has a thermal photon count statistics with autocorrelation $g_{\text{idler,auto}}^{(2)}=g_{\text{signal,auto}}^{(2)}=g_{\text{therm,auto}}^{(2)}\leq2$;
hence, according to Cauchy–Schwarz inequality the upper classical bound
on the second order cross-correlation function is $g_{\text{classical}}^{(2)}\leq\sqrt{g_{\text{signal,auto}}^{(2)}g_{\text{idler,auto}}^{(2)}}=2$.

\paragraph*{Schmidt number}

We estimate the number of modes by considering the theoretical prediction
for the biphoton wavefunction, given by Eq. (\ref{eq:psi}). We numerically
compute the wavefunction in a range of experimentally observed wavelengths
and transverse wavevectors. The obtained tensor is reshaped into a
two dimensional matrix with a single dimension corresponding to the
spectral and transverse coordinates of a single photon. Singular value
decomposition of the resulting matrix yields the Schmidt coefficients
$\{\lambda_{j}\}$. The Schmidt number (defined as per ref. \citep{Zielnicki2018}) is given by:
\begin{equation}
M=(\sum_{j=0}^{\infty}\lambda_{j}^{2})^{-1}\approx4.7,
\end{equation}
and corresponds to the approximate number of accessible entangled modes.

\paragraph*{Single-photon sensitivity of the custom CMOS camera}
The image intensifier (II) is operated in the Geiger mode (on/off). A photon striking the II leads to
the emission of a photoelectron (with quantum efficiency of $20\%)$
which is accelerated ($-200\;\text{V}$) or stopped ($+50\;\text{V}$)
with an electric potential controlled by the gating module. The accelerated
photoelectrons are multiplied via avalanche secondary emission in
a two-stage microchannel plate (MCP) giving $10^{6}$ output electrons
per photoelectron. The voltage across MCP is $1800\;\text{V}$. Electrons
leaving the MCP are further accelerated ($6000\;\text{V}$) and strike
a phosphor (P46) screen, leaving bright flashes. The phosphor screen
is imaged (magnification $M=0.44)$ with a relay lens onto the CMOS
sensor. We note that while we cross-correlate two regions for signal and idler
photons, respectively, an auto-correlation setup could be employed
with a single region and simple post-processing \citep{Lipka2018}.

\paragraph*{Setup efficiency}
We estimate the efficiency using a reference free method. Assuming a noiseless case, a single spatial mode and $R$ temporal modes per camera frame we have $\langle n_x \rangle= R \eta \chi ,\; x\in\lbrace i,s \rbrace$, $\langle n_i n_s \rangle = R \eta^2 \chi + \langle n_i \rangle \langle n_s \rangle$, where $\eta$ is the overall efficiency. Hence,
\begin{equation}
\eta = \frac{\langle n_i n_s\rangle - \langle n_i \rangle \langle n_s \rangle}{\sqrt{\langle n_i \rangle \langle n_s \rangle}}=\frac{g^{(2)}-1}{\sqrt{\langle n_i \rangle \langle n_s \rangle}}.
\end{equation}
Using $g^{(2)} = \max_{\lambda_+,k_+}[g^{(2)}(\lambda_+,k_+)]$ and global average numbers of photons $\langle n_i \rangle $, $\langle n_s \rangle$ we get $\eta\approx 4\%$. 
\end{document}